\documentclass[
 reprint,
superscriptaddress,
showpacs,preprintnumbers,
amsmath,amssymb,
aps,
pra,
]{revtex4-1}

\usepackage{graphicx}%
\usepackage{dcolumn}%
\usepackage{bm}
\usepackage{color} 
\usepackage{comment}
\usepackage{ulem}
\includecomment{comment}
\begin{document}


\title{Effect of optically-induced potential on the energy of trapped exciton-polaritons below the condensation threshold}

\author{M.~\surname{Pieczarka}}
 \affiliation{Nonlinear Physics Centre, Research School of Physics and Engineering,
The Australian National University, Canberra ACT 2601, Australia}
 \affiliation{ARC Centre of Excellence in Future Low-Energy Electronics Technologies}
 \affiliation{OSN, Department of Experimental Physics, Faculty of Fundamental Problems of Technology, Wroclaw University of Science and Technology, Wyb. Wyspia\'{n}skiego 27, 50-370 Wroclaw, Poland}

\author{M.~\surname{Boozarjmehr}}
\author{E.~\surname{Estrecho}}
 \affiliation{Nonlinear Physics Centre, Research School of Physics and Engineering,
The Australian National University, Canberra ACT 2601, Australia}
 \affiliation{ARC Centre of Excellence in Future Low-Energy Electronics Technologies}
\author{Y.~\surname{Yoon}}
 \affiliation{Department of Chemistry, Massachusetts Institute of Technology, Cambridge, MA 02139, USA}
\author{M.~\surname{Steger}}
 \affiliation{National Renewable Energy Lab, Golden, CO 80401, USA}
\author{K.~\surname{West}}
\author{L.~N.~\surname{Pfeiffer}}
 \affiliation{Department of Electrical Engineering, Princeton University, Princeton, NJ 08544, USA}
\author{K.~A.~\surname{Nelson}}
 \affiliation{Department of Chemistry, Massachusetts Institute of Technology, Cambridge, MA 02139, USA}
\author{D.~W.~\surname{Snoke}}
 \affiliation{Department of Physics and Astronomy, University of Pittsburgh, Pittsburgh, PA 15260, USA}
\author{A.~G.~\surname{Truscott}}
 \affiliation{Laser Physics Centre, Research School of Physics and Engineering,
The Australian National University, Canberra ACT 2601, Australia}
\author{E.~A.~\surname{Ostrovskaya}}
 \email{elena.ostrovskaya@anu.edu.au}
 \affiliation{Nonlinear Physics Centre, Research School of Physics and Engineering,
The Australian National University, Canberra ACT 2601, Australia}
 \affiliation{ARC Centre of Excellence in Future Low-Energy Electronics Technologies}
\date{\today}

\begin{abstract}

Exciton-polaritons (polaritons herein) offer a unique nonlinear platform for studies of collective macroscopic quantum phenomena in a solid state system. Shaping of polariton flow and polariton confinement via potential landscapes created by nonresonant optical pumping has gained considerable attention due to the degree of flexibility and control offered by optically-induced potentials. Recently, large density-dependent energy shifts (blueshifts) exhibited by optically trapped polaritons at low densities, below the bosonic condensation threshold, were interpreted as an evidence of strong polariton-polariton interactions [Nat.~Phys.~\textbf{13}, 870 (2017)]. In this work, we further investigate the origins of these blueshifts in optically-induced circular traps and present evidence of significant blueshift of the polariton energy due to reshaping of the optically-induced potential with laser pump power. Our work demonstrates strong influence of the effective potential formed by an optically-injected excitonic reservoir on the energy blueshifts observed below and up to the polariton condensation threshold and suggests that the observed blueshifts arise due to interaction of polaritons with the excitonic reservoir, rather than due to polariton-polariton interaction.

\end{abstract}

\maketitle
\section{Introduction}
Studies of exciton-polaritons (or simply polaritons) in quantum wells embedded into semiconductor microcavities have developed into an active research field driven by the ability to observe condensation \cite{Deng2002,Kasprzak2006,Balili2007, Cilibrizzi2014,Wertz2010,Christopoulos2007,Deng2010,Carusotto2013} 
and superfluidity \cite{Lerario2017,Amo2009,Nardin2011} of these quasiparticles on a well-developed solid-state platform. The effectively repulsive interaction of polaritons stemming from the Coulomb interaction of their excitonic constituents \cite{Tassone1999, Vladimirova2010, Takemura2014} not only assists the condensation via stimulated bosonic scattering, but also leads to a wealth of nonlinear mean-field effects observed at higher densities \cite{Rodriguez2016}. The precise value of the strength of the polariton-polariton interaction has recently become a subject of controversy, as a recent measurement performed well below the condensation threshold, i.e. in the low polariton density regime \cite{Sun2017}, resulted in a quantity which is at least two orders of magnitude larger than that previously accepted by the exciton-polariton community \citep{Ferrier2011, Vladimirova2010, Tassone1999}. The importance of this claim cannot be underestimated as it implies that the polaritons are, in fact, strongly interacting particles that can be naturally driven to strongly correlated quantum phases even at a very low density.

The above-mentioned claim is based on the measurement of the upward energy shift (blueshift) of the low-density, below the condensation threshold polaritons accumulating in an optically-induced circular trap. The latter is defined by an annular potential barrier created by a ring-shaped off-resonant optical pump which photoinjects an incoherent reservoir of highly energetic excitonic quasiparticles \cite{Cristofolini2013,Askitopoulos2013,SunBEC2017,Sun2018,Ohadi2015,Dreismann2014}. The reservoir feeds polaritons and confines them through repulsive interactions. The critical assumption made in \cite{Sun2017} is that the excitonic reservoir is spatially localised at the position of the maximum intensity of the optical pump due to its large effective mass and low mobility, and that its effect is negligible in the middle of the resulting circular well trap. The blueshift measured in \cite{Sun2017} at zero momentum (kinetic energy) is therefore attributed purely to the polariton-polariton interaction energy, and appears to be anomalously large given the very low densities. 

In this work, we report a detailed investigation of the below-condensation behaviour of polaritons in optically-induced circular potential wells of various diameters and different fractions of photon and exciton in the polariton quasiparticle. Our experiment and modeling suggest that the significant blueshift of the polariton energy at low densities originates from the interactions of the polaritons with the reservoir, rather than solely from polariton-polariton interactions. At very low densities and pump powers, photon-like polaritons experience a strong quantum confinement effect \cite{Cristofolini2018,Schneider2017,ElDaif2006}, which results in quantisation of energy levels which dominates the blueshift as the shape of the trapping barrier changes with the increasing pump power. The quantum confinement effect for exciton-like polaritons in large-area traps is negligible, but the significant shifts of the lowest polariton energy at zero momentum are caused by the rising bottom of the potential trap due to the spreading of the excitonic reservoir. Additionally, we describe the challenges of the methodology of the blueshift measurements for polariton photoluminescence below condensation threshold. We conclude that the energy blueshifts of optically confined polaritons measured at low densities in momentum space cannot be used to determine the polariton-polariton interaction strength. 

The manuscript is structured as follows. In Sec.~\ref{sec:methodology} we describe the sample, the experimental setup, and methodology of the measurement. Section \ref{sec:results} contains the main findings of our work and comparison with the previously reported results, and is followed by the discussion in Sec. \ref{sec:discussion}. Section \ref{sec:conclusions} summarises and concludes the work.

\section{\label{sec:methodology}Methodology}
\subsection{Experimental details}
We study a high-quality GaAs/AlAs microcavity sample with $12$ embedded 7-nm quantum wells and distributed Bragg reflector (DBR) mirrors consisting of a large number (32 top and 40 bottom) of layer pairs  \cite{Nelsen2013,Sun2017,Myers2018}. The sample design and its specific properties, such as a long polariton lifetime and a large position-dependent gradient of the cavity photon energy $E_{ph}$ due to a wedge in the cavity thickness \cite{Nelsen2013,Sun2017}, are similar to the sample used in the recent studies of the ground state blueshift of optically trapped polaritons \cite{Sun2017}. 

\begin{figure}
	\includegraphics{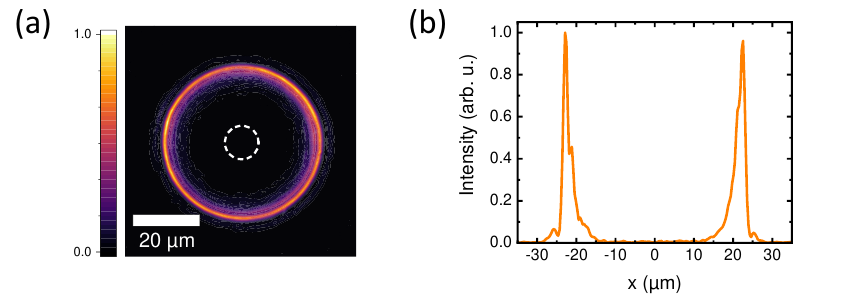}
\caption{\label{fig1} (a) Image of a reflected optical excitation intensity from a sample surface presenting the shape of the pump distribution for a ring of $45$ $\mu$m in diameter. White dashed circle indicates the position and shape of the spatial filter. (b) A cross-section through the middle of the intensity ring showing a clean centre of the excitation pattern (the scattered light in the middle is about $< 0.8$\% of the light intensity at the ring position).}
\end{figure}

The experimental setup is similar to the one used in our previous works \cite{Gao2018,Dall2014}. The sample is kept in a cryostat ensuring a temperature of around $7$ K. To generate a circular well potential, a nonresonant continuous-wave (\textit{cw}) laser, chopped by an acousto-optic modulator at 10~kHz and 5\% duty cycle, is shaped to an annular intensity distribution on the sample surface, as shown in Fig.~\ref{fig1}. This is achieved by introducing an axicon between a pair of confocal lenses in the path of the laser beam \cite{Manek1998}. This technique enables generation of annular intensity patterns of various diameters $25$-$70$~$\mu$m in the current configuration, annulus thickness of $3$-$4$~$\mu$m, and a clean interior of the trap, see Fig. \ref{fig1}(b). The laser wavelength is tuned to the second reflectivity minimum of the microcavity to create free electron-hole pairs in the quantum wells at the pump position, which relax down to form highly energetic excitonic quasiparticles forming the reservoir. Naturally, the reservoir particles exist in the vicinity of the laser pump and hence follow its ring-shape distribution. Polaritons are then created from the reservoir where a significant proportion of polaritons is pushed towards the centre of the ring due to its repulsive interaction with the excitonic reservoir particles forming a trap \cite{Cristofolini2013,Askitopoulos2013}. 

\subsection{Polariton energy in an optically-induced trap} 

\begin{figure} 
	\includegraphics[scale=0.125]{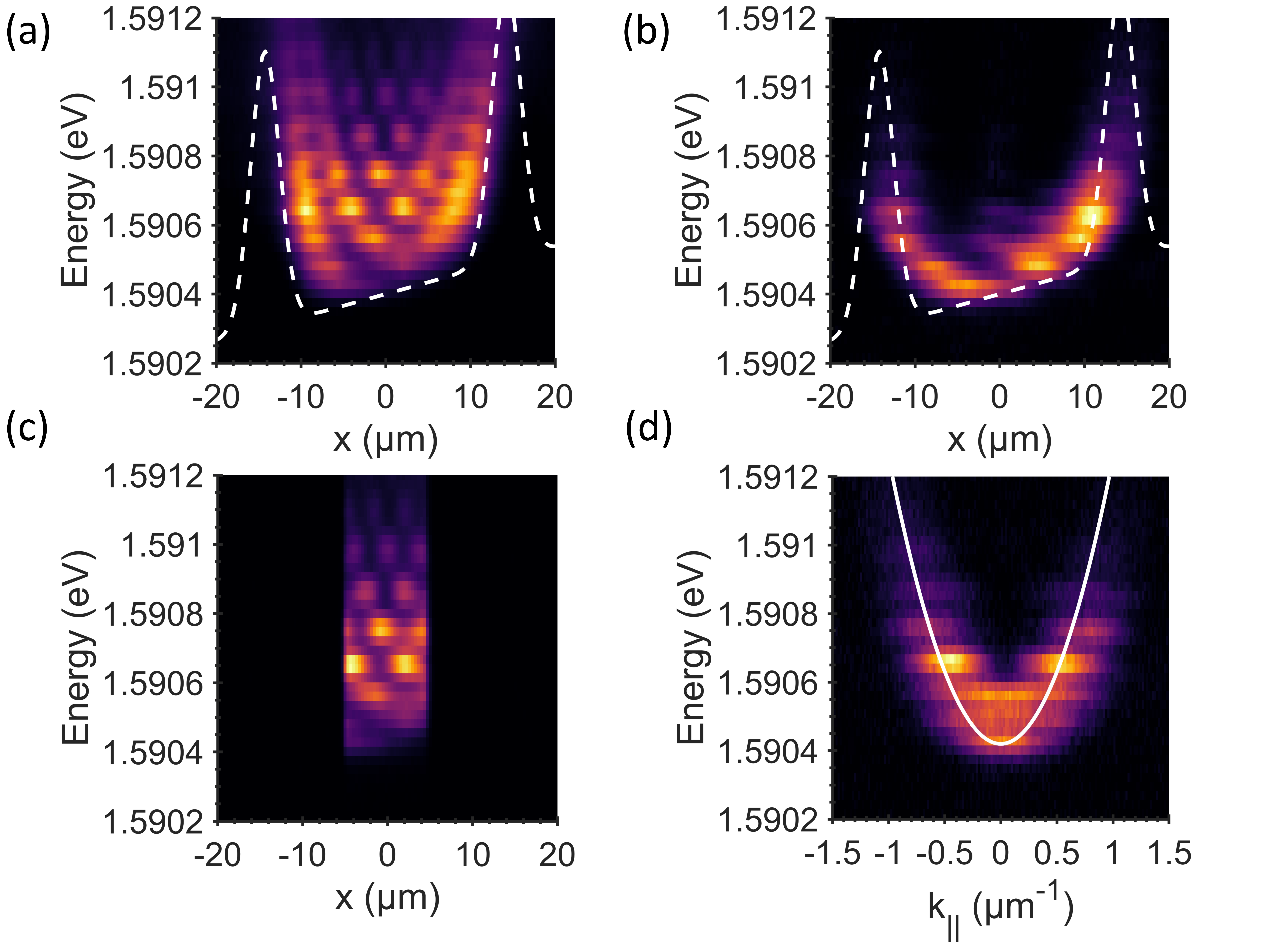}
\caption{\label{fig2} Exemplary spectra recorded in the experiment in the case of photonic detuning $\Delta = -12$ meV and small trap $D = 29~\mu$m at the pump power $17$ mW. (a) Full real space spectrum of confined energy states. (b) ${\bf k_{\parallel}}=0$ filtered real space spectrum showing the contribution of the ground state and the classical turning points of excited states. (c) Real space spectrum with imposed spatial filter in the centre. (d) Resulting dispersion of the filtered area from (c). One can easily match the corresponding quantised states contributing to the ${\bf k_{\parallel}} = 0$ signal. Dashed lines in (a) and (b) correspond to the deduced approximation of an effective potential. Solid line in (d) depicts the theoretical polariton dispersion.}
\end{figure}

In the work of Ref. \cite{Sun2017}, the linear dependence of the lowest energy of the trapped polaritons at zero in-plane momentum (${\bf k}_{||}=0$) on the polariton density was interpreted as a consequence of the repulsive interactions among the increasing number of polaritons within the trap. This interpretation is based on the expression for the mean-field energy of the polaritons due to polariton-polariton interactions, $E = gn$, where $g$ is the interaction constant and $n$ is the polariton density. The value of $g$, extracted as the value of the linear slope ($E/n$) below threshold, is around two orders of magnitude larger than the commonly accepted theoretical estimation of the interaction strength \cite{Tassone1999}. However, the mean-field, i.e. density-dependent, contribution to the polariton energy shift is likely to be negligible below the condensation threshold, where the polariton density is very low. Larger contributions are expected due to the repulsive interactions between the polaritons and reservoir particles, which create a local blueshift (effective potential) proportional to the reservoir density $n_{R}$: $V_{R}({\bf r})=g_{R}n_{R}({\bf r})$, where $g_{R}$ is the strength of interaction between polaritons and the photo-injected excitonic reservoir \cite{Wertz2010,Ferrier2011}. The total (potential and kinetic) polariton energy in the low-density limit, in the presence of the reservoir can be calculated from the Schr\"{o}dinger equation:
\begin{equation}
-\frac{\hbar^2}{2m^{*}}\nabla^2 \psi+V_{\rm eff}({\bf r})\psi=E\psi,
\label{SE}
\end{equation}
where $\psi$ is the single-polariton wavefunction, $m^*$ is the effective mass of the polariton, $V_{\rm eff}({\bf r})=V_R({\bf r})+E^{0}_{LP}({\bf r})$, and $E^{0}_{LP}({\bf r})$ is the minimum of the single particle lower polariton energy $E^0_{LP}=E_{LP}({\bf k_{\parallel}}=0)$ in the absence of the reservoir. Any change of the reservoir density will lead to a change of $V_{\rm eff}$ and to a corresponding shift of the polariton energy $E$. According to conventional understanding, the interaction strengths are $g_{R}\sim |X|^2g_{X}$, and $g\sim |X|^4g_{X}$, where $|X|\leq 1$ is the excitonic Hopfield coefficient, and $g_X$ is the strength of exciton-exciton interactions \cite{Deveaud-Pledran2016,Carusotto2013,Walker2017}. Below condensation threshold the reservoir density $n_R >> n$, therefore, the effects of polariton-reservoir $g_R n_R$ interactions may exceed those of polariton-polariton $g n$ interaction even by two orders of magnitude. In our experiments, by exploiting the large wedge in the microcavity \cite{Sun2017} and hence the large gradient of the cavity mode energies, we are able to explore a wide range of exciton-photon detuning values, $\Delta=E_{ph}-E_X$, corresponding to a range of different excitonic Hopfield coefficients $|X|^2$, hence polaritons experiencing different $g_R$. In addition, the wedge leads to an effective potential gradient arising from the spatial dependence of the minimum of the polariton energy: $E^0_{LP}=E^0_{LP}({\bf r})$, as seen in Fig. \ref{fig2}(a,b). 

In order to understand the influence of the effective potential $V_{\rm eff}({\bf r})$ on the polariton energy shifts in the trap (exemplary real space spectrum is presented in Fig. \ref{fig2}(a)), we perform momentum ($k$) space spectroscopy of the cavity photoluminescence and employ an integration technique similar to that used in \cite{Sun2017} (details are given in {\color{blue}Appendix A}). We introduce an iris as a spatial filter ($\sim 10$ $\mu$m diameter in the real space image plane) as shown in Fig.~\ref{fig1}, to eliminate the signal from the high-energy polaritons in the pump region and probe only polaritons in the centre of the trap, see Fig.~\ref{fig2}(c). The resulting polariton dispersion is shown in Fig.~\ref{fig2}(d), where one can isolate the spectrum at zero in-plane momentum ${\bf k}_{||}=0$. Contrary to the assumption of Ref.~\cite{Sun2017}, the ${\bf k}_{\parallel}=0$ state is not necessarily the polariton ground state energy of the trap. This is due to the strong local energy gradient in the sample, arising from the wedge of the microcavity, which results in an effective ``triangular'' potential well (see Fig.~\ref{fig2}) that can lead to non-ground states of the potential well to have ${\bf k_{\parallel}}=0$ contributions. To distinguish the contributions of different energy states to the total signal at ${\bf k}_{||}=0$, one can introduce a filter in the conjugate plane (${\bf k}$-space) of the optical setup, filtering only the ${\bf k}_{||}\approx 0$ from the real space image (see {\color{blue}Appendix B}). As shown in Fig.~\ref{fig2}(b), the ${\bf k}_{||}\approx 0$ signal has contributions from different energy states corresponding to the classical turning points (zero kinetic energy) of the confined states in the trap, therefore depicting the approximate spatial shape of the reservoir-induced potential energy landscape $V_{\rm eff}({\bf r})$.

\begin{figure}
	\includegraphics[scale=0.09375]{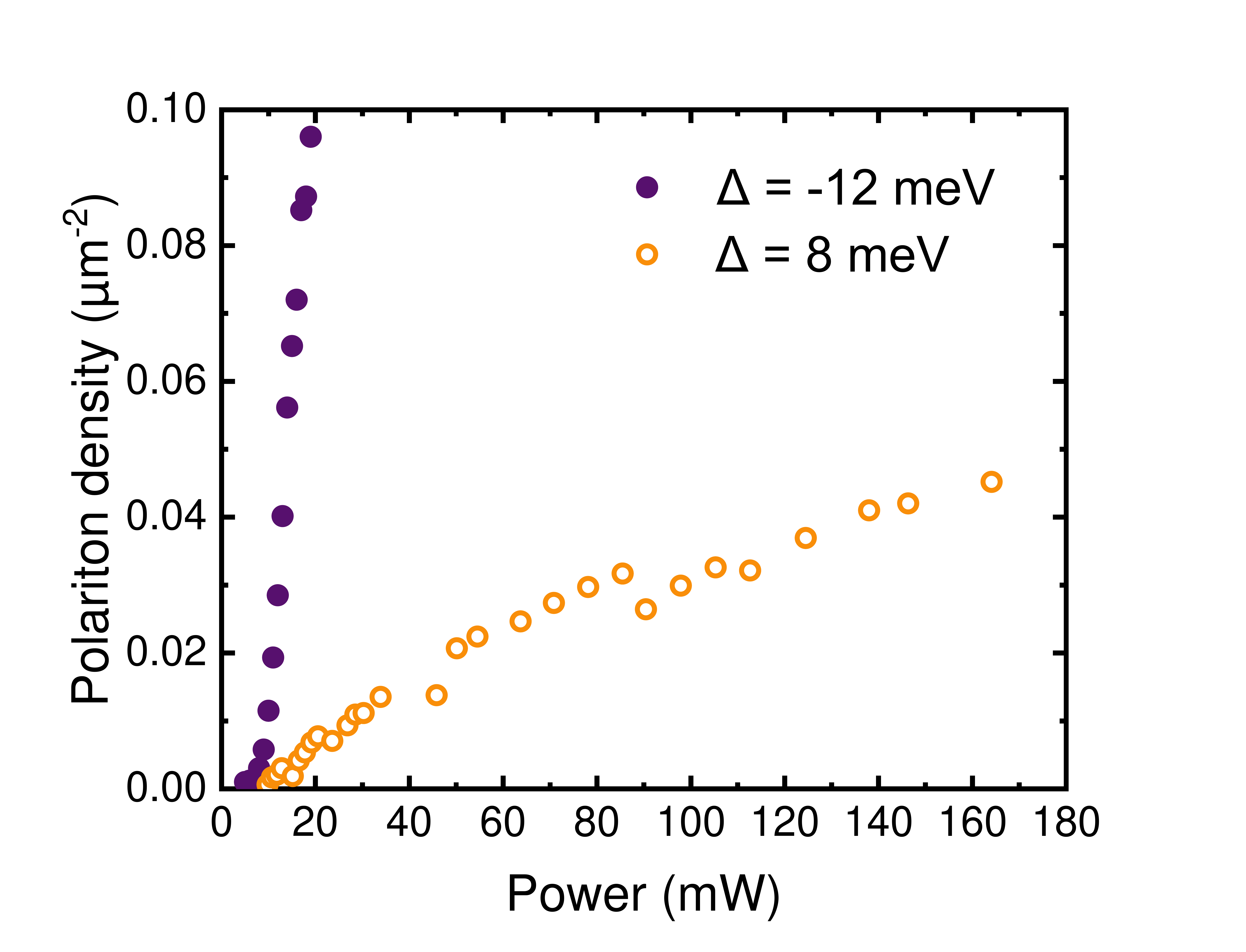}
\caption{\label{fig3} Power-dependent density in the middle of large traps of $D = 45~\mu$m in the photonic and excitonic detuning case showing the enormous difference in the efficiency of polariton generation inside the trap.}
\end{figure}

\section{\label{sec:results} Results}

The density of polaritons in the centre of the trap grows with the pump power and so is the reservoir density in the vicinity of the annular laser pump. We can distinguish between two possible effects of the reservoir-induced potential on the polariton energy. One is the change of the barrier height and the trap area driven by the pump power, which will have a profound effect on the energy eigenstate, $E$, in the effective potential $V_{\rm eff}$ as found from Eq.~(\ref{SE}).  Another effect is the buildup of the reservoir near the centre of the trap, which could also lead to significant blueshift  of the zero-point energy, since in the absence of the reservoir density in the middle of the ring, $V^0_{\rm eff}=E^0_{LP}$. Both effects will lead to significant blueshift of the confined polariton energy in the case when the size of the trap is small enough and the energy levels are strongly quantised. The latter effect will strongly influence the lowest polariton energy even in the case when there is no quantum confinement effect, i.e. when the polaritons can be considered as a classical gas with a continuous spectrum.

In what follows, we analyse two extreme cases of exciton-polariton detuning: highly photonic $\Delta = -12$ meV, corresponding to $|X({\bf k_{\parallel}}=0)|^2 \approx 0.21$ and highly excitonic $\Delta = +8$ meV, corresponding to $|X({\bf k_{\parallel}}=0)|^2 \approx 0.73$. Photon-like, low-mass polaritons quickly fill the middle of the trap as they propagate ballistically away from the excitation region with large velocities $v\propto \sqrt{V^{\rm max}_{\rm eff}/m^*}$ \cite{Kasprzak2008,Sun2018,Wouters2008}. On the other hand, exciton-like polaritons at positive detunings have a larger effective mass and are subject to more efficient phonon-assisted energy relaxation \citep{Kasprzak2008,Deng2006,SunBEC2017}, thus tending to accumulate in the area of the potential barrier defined by the annular pump. Hence, at different detunings, different laser pump powers, i.e., different barrier heights, $V^{\rm max}_{\rm eff}$, are required to achieve the same polariton concentration inside the trap. The magnitude of this difference is illustrated in Fig. \ref{fig3}, where we compare the integrated polariton densities at given pumping powers inside a large trap of $45$ $\mu$m diameter. One can observe a significant difference in the polariton generation yield inside the trap below condensation threshold, where polaritons at the photonic detuning are created around $20$ times more efficiently than at the excitonic detuning. This means that, to achieve a comparable density of photonic and excitonic polaritons in the trap, one has to reach at least an order of magnitude larger reservoir density in the excitonic case, which has important implications for the interpretation of the experimental results in these two extreme cases.

\subsection{Quantum confinement effect in the photonic detuning regime} 

\begin{figure}
	\includegraphics{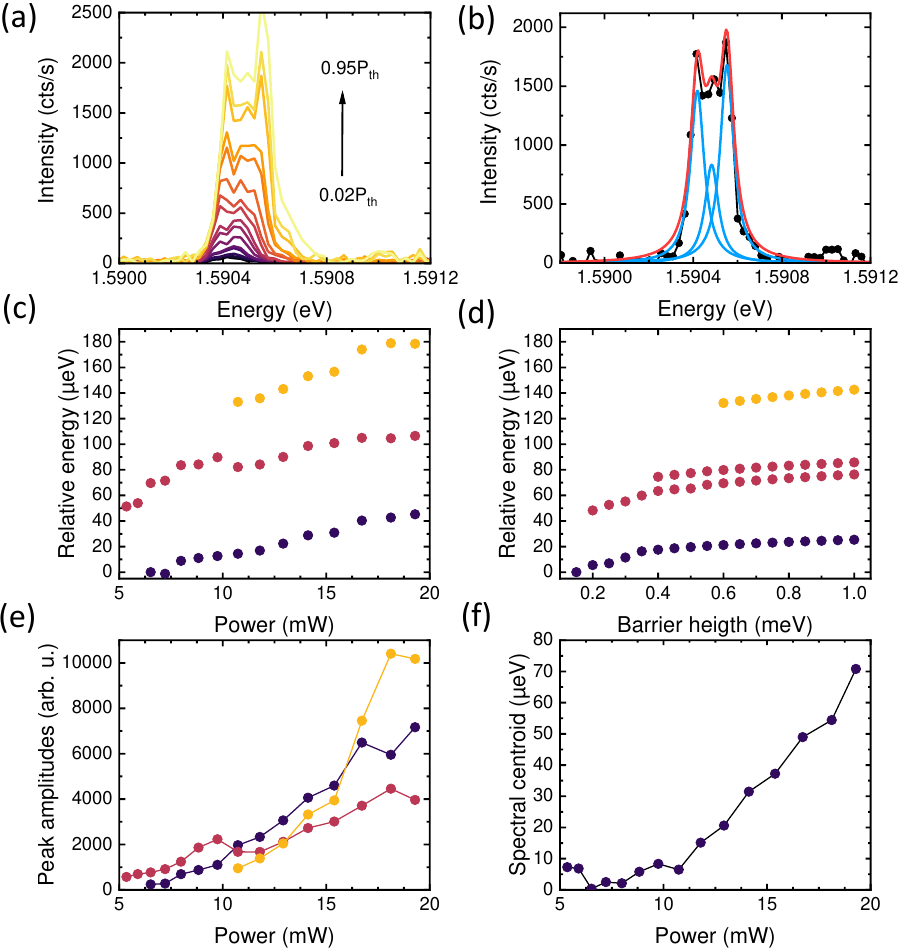}
\caption{\label{fig4} Analysis of the photonic detuning case in a small trap of $D = 29~\mu$m, where $P_{th} \approx 20$~mW. (a) Series of power dependent spectra of ${\bf k_{\parallel}}=0$ in the range from $P=0.02P_{th}$ to $P=0.95P_{th}$. (b) Example of a fit for spectrum at $P=0.84P_{th}$. (c) Power dependency of extracted energies of the states contributing to the ${\bf k_{\parallel}}=0$ spectrum. (d) Lowest eigenstates calculated for a trap of different barrier heights. (e) Amplitudes of extracted peaks. (f) Spectral centroid of the ${\bf k_{\parallel}}=0$ experimental data representing the net blueshift.}
\end{figure}

Firstly, we describe the experimental data taken for a smaller trap of diameter $D = 29$~$\mu$m, where the quantum confinement effect is more pronounced due to the low effective mass and the energies of the quantised eigenstates are clearly resolvable. As described in the previous section, we analyse the ${\bf k_{\parallel}}=0$ signal taken from the central area of the trap by using a spatial filter with a diameter of about 10~$\mu$m. The power-dependent spectra below the condensation threshold are summarised in Fig.~\ref{fig4}(a). One can observe that the measured spectral line profile does not correspond to a single peak shape, neither Lorentzian nor Voigt. As also seen from Fig.~\ref{fig2}, the measured ${\bf k_{\parallel}}=0$ spectrum is shaped by a contribution of several quantised energy states within the field of view defined by the imposed spatial filter. The origin of these contributions is the ${\bf k_{\parallel}}=0$ component of each of the confined states corresponding to the classical turning points of the polaritons in the effective potential. 

In the particular case presented in Fig.~\ref{fig4}, the analysis of the line profile is performed by fitting the ${\bf k_{\parallel}}=0$ signal with three Lorentzian lines, assuming a resolution limited broadening of each component (experimental setup resolution is about $75 \pm 5$~$\mu$eV). An example of such fitting is presented in Fig.~\ref{fig4}(b), and the extracted energies are plotted in Fig.~\ref{fig4}(c). All states experience a pump-dependent blueshift, which saturates to the value of about $30$-$40$ $\mu$eV for each line. The recorded energy shifts are in a good agreement with the numerical modelling, as seen in Fig. \ref{fig4}(d). To obtain the numerical values, we solved the two-dimentional Schr\"{o}dinger equation, Eq.~(\ref{SE}), taking into account the polariton effective mass and the position-dependent energy gradient at the particular value of detuning, and assuming 4 $\mu$m thick (FWHM) Gaussian-shaped walls of the ring potential. The potential height was varied while keeping the wall thickness constant, which emulates the growth of the reservoir density at the position of the pump. The calculated lowest-lying energies correspond well to the experimental points, as seen in Figs.~\ref{fig4}(c,d), where one of the experimentally determined states is represented by a doublet of near-degenerate states in the numerical simulation. 

We also integrated the experimental data and extracted the spectral centroid of the multiple peaks, where the result of such analysis is depicted in Fig. \ref{fig4}(f). The value of this shift is of the same order as observed previously \cite{Sun2017}. The origin of the net shift is the blueshift due to quantum confinement arising from the pump power-dependent potential reshaping combined with redistribution of the ${\bf k_{\parallel}}=0$ contributions from the low-energy confined states. The redistribution of occupancies is due to the lack of energy relaxation and thermalisation in polariton gases at negative detunings \cite{SunBEC2017}. The combination of these effects  results in a change of the line profile with rising pump power which can be misinterpreted as an increase in blueshift with growing polariton density. This conclusion is supported by analysis of separate amplitudes of the individual energy peaks corresponding to the trapped states extracted from the experimental data, as presented in Fig.~\ref{fig4}(e). One can observe the interplay of contributions between the three energy states with changing pump power. 

\begin{figure}
	\includegraphics[scale=0.125]{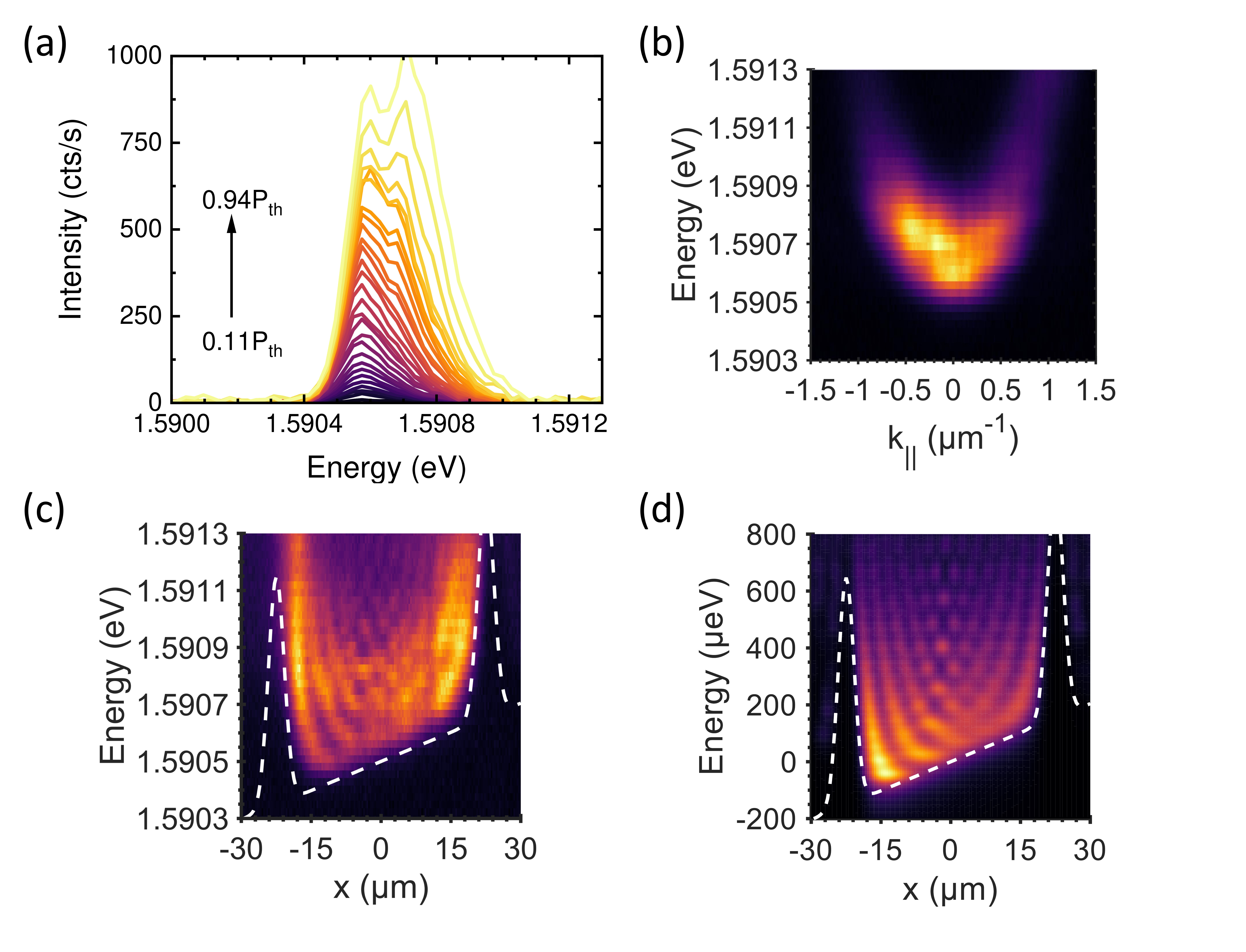}
\caption{\label{fig5} Results of power-dependent blueshifts of the photonic detuning case in a large trap $D = 45~\mu$m, where $P_{th} \approx 36$~mW. (a) Series of power dependent spectra of ${\bf k_{\parallel}}=0$ in the range from $P=0.11P_{th}$ to $P=0.94P_{th}$. (b) Far-field dispersion of measured polariton emission at $P=0.77P_{th}$. (c) Real space experimental spectrum of the confined states in the trap at $P=0.77P_{th}$. (d) Simulation of a polariton luminescence, based on 2D Schr\"{o}dinger equation and assuming thermal population of the states.}
\end{figure}

Next, we examine the case of a similar photonic detuning, but in a larger trap of $45$ $\mu$m in diameter, where the quantum confinement is expected to have a negligible contribution to the blueshift. The ${\bf k_{\parallel}}=0$ spectra are presented in Fig.~\ref{fig5}(a) and display similarity to those observed in a smaller trap, i.e. the spectral line shape is a result of an overlap of many closely spaced spectral peaks. Once again, the net blueshift is visible and originates from the quantum confinement effect and the reshaping of the spectral line. Performing similar numerical simulation as in the previous case, we find that the ${\bf k_{\parallel}}=0$ spectrum is composed of at least 6 spectral lines originating from the confined states in the field of view, making the extraction of the constituent energies from the experimental data impossible to perform reliably. However, the quantum confinement is still apparent in this trap, as confirmed by the comparison of the experimental real space spectrum with the simulated one in Fig.~\ref{fig5}(c) and (d), respectively, which shows the specific pattern of the quantised states inside the trap. In this larger area trap, the spacing between the states is smaller, therefore the ${\bf k_{\parallel}}=0$ line profile is much smoother in comparison to the small trap, and the confined states are hardly visible in the spatially filtered far-field ($k$-space) spectrum, as shown in Fig.~\ref{fig5} (b). In this figure, one can see an additional effect contributing to the ${\bf k_{\parallel}}=0$ signal, namely some portion of the ${\bf k_{\parallel}} \neq 0$ states may add to this spectrum as each ${\bf k_{\parallel}}$ state is significantly broadened due to the real-space filtering with the filter size of 12~$\mu$m in diameter (see discussion in section~\ref{sec:discussion}). As a result, one observes both the broadening and the net shift of the ${\bf k_{\parallel}} \neq 0$ spectrum.

\subsection{Trap reshaping and polariton energy in the excitonic detuning regime} 

In the  regime of exciton-like polaritons (large positive detuning) in large-area traps, the quantum confinement effect is negligible due to the decreasing level separation $\propto 1/(m^*D^2)$. The effect of the cavity wedge on the trap shape is also less pronounced at large positive detuning, which results in a potential with a nearly ``flat" bottom. As the effects of confinement and associated level separation become weaker, the spectrum approaches a continuum of states, which contribute to a large uncertainty in determining the energy of the ground state resulting from experimental methodology described in previous sections. Moreover, the polariton linewidth approaches the QW excitonic one \cite{Savona1997}, further broadening the states.

\begin{figure}
	\includegraphics[scale=0.125]{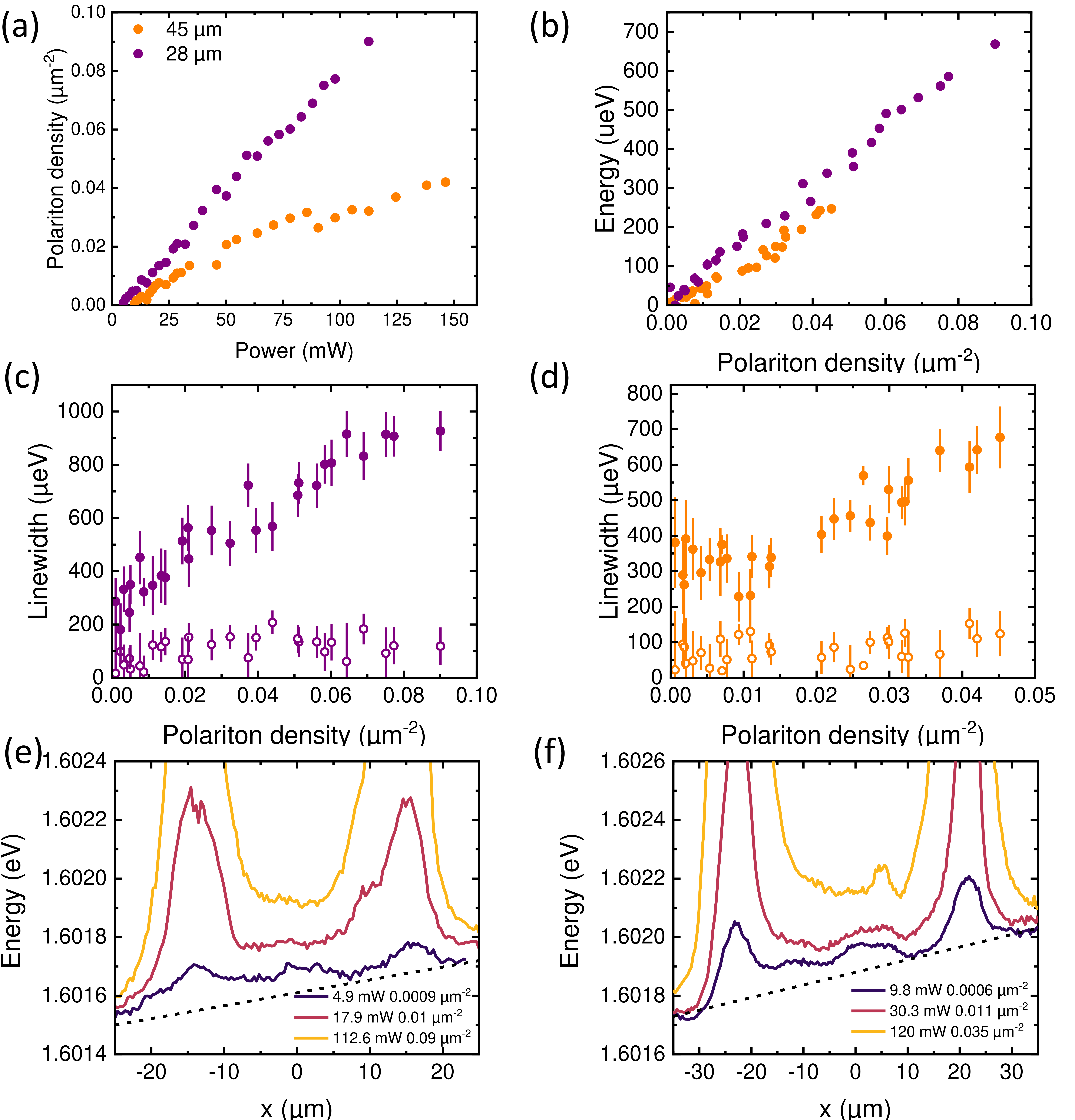}
\caption{\label{fig6} Summary of experimental data analysis for excitonic detuning $\Delta\approx +8$ meV and two trap diameters $D = 28~\mu$m (purple) and $D = 45~\mu$m (orange). (a) Power dependent density measurements. (b) ${\bf k_{\parallel}}=0$ energy shifts versus polariton density. Dependency of homogeneous (closed) and inhomogeneous (open circles) linewidths on the measured polariton density at $D = 28~\mu$m (c) and $D = 45~\mu$m (d) traps. Extracted potential energy shapes (e) $D = 28~\mu$m and (f) $D = 45~\mu$m. Dashed line indicates the extracted local ground state energy slope.}
\end{figure}

As in the photonic case, we investigate two different trap diameters $D = 29~\mu$m and $D = 45~\mu$m and the results of our analysis are summarised in Fig. \ref{fig6}. One can observe a difference in power dependent polariton densities for different trap diameters, Fig. \ref{fig6}(a), where polaritons accumulate in the middle of the trap more efficiently in a smaller trap case. Additionally, the density increases linearly with the pump power, whereas in the case of a large trap it is nonlinear and saturates indicating inefficient polariton generation. This effect is also due to the fact that in the case of the smaller trap, the pump density is larger, therefore excitation of carriers in the sample is more efficient. Despite the differences, the density dependent blueshifts of the ${\bf k}_{\parallel} = 0$ state show a similar linear behaviour, see Fig. \ref{fig6}(b), with the slopes of about $7.5$ meV$/\mu$m$^{-2}$ for the small trap and $5$ meV$/\mu$m$^{-2}$ for the large one, being of the same order of magnitude as the values obtained in \citep{Sun2017} for similar detuning.  

Due to fact that the ${\bf k}_{\parallel} = 0$ line profile consists of a continuum of a broadened states in the excitonic case, it is possible to analyse its behaviour by fitting the peak with a Voigt lineshape, allowing for extraction of homogeneous and inhomogeneous contributions to the spectrum. The extracted values of the broadening are consistent with the previous observations \cite{Sun2017}, where the inhomogeneous broadening is roughly constant (about 100-150 $\mu$eV) and the homogeneous broadening increases with polariton density, hence larger pumping powers, see Figs. \ref{fig6}(c) and (d). The constant value of the inhomogeneous broadening represents a finite set of states probed in the experiment, which are located along the slightly tilted bottom of the trap. The strong increase of the homogeneous linewidth broadening was previously interpreted as a consequence of polariton-polariton interactions \cite{Sun2017}. However, the same effect is expected from polariton-reservoir interactions \cite{Askitopoulos2013}, which are much stronger.

As described at the beginning of Sec. \ref{sec:results}, one has to pump strongly to obtain larger densities of highly excitonic polaritons inside the trap, thus unavoidably creating very large densities of the incoherent reservoir near the pump region. As a consequence, one can expect the optically-induced potential to change significantly with high pumping powers and the reservoir to be pushed into the middle of the trap. To verify this hypothesis, we performed measurements of the density-dependent trap shape change. This was done by introducing a k-space filter of approx 0.7-0.8 $\mu$m$^{-1}$ in diameter effectively filtering out the higher energy and high-${\bf k_{\parallel}}$ states and imaging the near-zero kinetic energy contribution, see Sec. \ref{sec:methodology} and {\color{blue}Appendix B}. In this case, the quantum confinement effect is negligible, therefore the resulting spectrum of low density polaritons follows the local effective potential $V_{\rm eff}({\bf r})$. The extracted trap shapes are presented in Figs.~\ref{fig6}(e) and (f). One can observe that the trap bottom does not follow the extracted sloped line of $E^0_{LP}({\bf r},{\bf k_{\parallel}=0})$ and tends to flatten with increasing injected carrier density, both for large and small trap case. This suggests, that reservoir particles accumulate inside the optical trap and the energy shift of the trap bottom becomes responsible for the energy shift of polaritons. Additionally, the trap diameter shrinks with increasing pump power, where the high density reservoir interactions at the annulus broadens the effective width of potential barriers, which might contribute to the effective shift of the polariton energy inside the trap.

\section{\label{sec:discussion}Discussion} 

Our results highlight the crucial difficulty in extracting the energy shifts of polaritons below the condensation threshold, where the spectrum consists of many overlapping energy states in the photonic detuning case, and approaches continuum of states in the excitonic detuning case. The detected signal in $k$-space is influenced by not only reshaping of the trapping potential, but also by the redistribution of occupancy of states contributing to the ${\bf k}_{\parallel} = 0$ signal and the experimental method of the signal filtering. Near-field (real space) and far-field ($k$-space) spectra are related to each other via spatial Fourier transform. Introducing a spatial filter to one of the conjugate planes produces broadening of the result in the other plane. Spatial and $k$-space resolution limits are related to each other as $\Delta x = 2\pi/\Delta k$, which implies that the introduction of too narrow filters in one plane violates the resolution limits. As an example, $\Delta x = 12~\mu m$ results in $\Delta k = 0.52~\mu m^{-1}$, which means that the ${\bf k}_{\parallel}\neq 0$ of higher order states can overlap with the ${\bf k}_{\parallel}\approx 0$ signal and this effect can contribute to the broadening and the spectral line shape of the resulting peak \cite{Sun2017}, as can be seen in Figs. \ref{fig5}(a,b). This also holds if one introduces too narrow $k$-space filtering thus degrading the spatial resolution, where spectrum at a given position is composed of overlapped adjacent ${\bf k}_{\parallel}\approx 0$ states. Thus, the analysed signal in both methods can have non-negligible contribution of the excited energy states in the recorded signal.

Regardless of the above constraints, both in the photonic and excitonic detuning regimes, our measurement shows significant reshaping of the reservoir-induced trap that takes place at large pumping powers. At photonic detunings this reshaping generates shifts due to the quantum confinement effect and lack of relaxation between the confined energy states. In the case of excitonic detunings, which require large pump powers to detect any polaritons in the middle of the trap, such reshaping includes broadening of the trap walls and resulting buildup of the reservoir particles in the middle of the ring, which means a significant contribution of the reservoir density, $V_{R}$ to $V_{\rm eff}$ in the trap centre. Additionally, in the excitonic regime, the linewidths of the individual energy states are significantly broadened, making it difficult do isolate the true ground state of the trap. Furthermore, the spatially resolved potential energy measurements $E({\bf r})$ (at ${\bf k}_{\parallel} \approx 0$) at high pump powers presented in Fig. S11 of Ref.~\cite{Sun2017} and in Fig. \ref{fig6} of this work indicate that, in this regime, the magnitude of the shift of the trap bottom is responsible for the overall blueshift of the lowest ${\bf k}_{\parallel} \approx 0$ energy of the polaritons in the detection window. 

In the intermediate case, of the near-zero exciton-photon detuning $\Delta \approx 0$, there is an interplay of both mechanisms governing the energy shifts of low density polaritons (not shown here). The effect of quantum confinement is not well resolvable due to the much larger linewidths and smaller separation compared to the photonic case. Nevertheless, the buildup of reservoir inside the trap can also account for the anomalously large energy shifts. The blueshifts are smaller in comparison to the strongly excitonic case presented above, as the laser pump power required for obtaining large population of polaritons in the middle are lower \cite{Sun2017}, thus the reservoir density is lower as well.

The physical reason for the significant reservoir buildup in the centre of the trap should be carefully investigated in further experiments. One possibility is that, while the diffusion of the heavy excitons in these samples is considered to be small (of the order of $\sim 1$ $\mu$m), the reservoir composed of mobile, high-momentum bottleneck polaritons could result in a significant diffusion into the centre of the trap. This possibility is supported by the recent study \cite{Myers2018}, which found the diffusion lengths of the highly excitonic polaritons at high ${\bf k}_\parallel$ to be greater than $20$ $\mu$m. It is also supported by the observation here and in Ref.~\cite{Sun2017} of large homogeneously broadened linewidths, which could originate from interaction with the reservoir particles. Furthermore, nonresonantly excited quantum well excitons at high densities can also have large transport lengths of the order of 10 $\mu$m and atypical distribution profiles, being a possible source of a trap bottom shift at high pumping powers \cite{Smith1989}.

\section{\label{sec:conclusions}Conclusions}
We investigated optically confined exciton-polaritons in the low-density regime, where the density-dependent polariton-polariton interaction energy is negligible. Power and density-dependent energy shifts observed in our experiment are comparable with the data presented in \cite{Sun2017}. A detailed analysis of polariton energies at different detunings and trap diameters shows that the effect of interactions between the polaritons and the optically-injected excitonic reservoir can account for anomalously large blueshift of polariton energy below the condensation threshold. 

Our analysis shows that the dominant effect that results in the blueshifts of the ${\bf k}_{\parallel} = 0$ spectrum in a pump-induced effective potential trap in the photonic regime is the quantum confinement coupled to the pump power-dependent potential wall height. In large-area traps and for excitonic detuning the expected confinement-induced blueshift is small compared to that observed in the experiment. Since, in this regime, the zero-point energy coincides with the bottom of the trap, we suggest that buildup of the reservoir particles in the middle of the trap leads to the anomalously large blueshifts reported in \cite{Sun2017} and observed in our experiments. On the basis of our observations, we conclude that the optically-confined polaritons are strongly influenced by the interaction with the excitonic reservoir, and the blueshift of optically-trapped polaritons below the condensation threshold cannot be unambiguously related to the polariton-polariton interaction strength.

We note, however, that accurate measurements of the polariton-polariton interaction strength can still be performed above the condensation threshold, where the macroscopic occupation of a single particle state, being ground state of the system, can lead to large polariton densities and significant density-dependent blueshifts \cite{Estrecho22018}.

\section*{Appendix A: momentum-space integration}

The essential part of the performed work is to carefully calibrate the experimental setup to determine the blueshift of the ground state and the polariton density generated inside the circular trap. Firstly, we calibrate the collection efficiency of our setup, to determine how many photons emitted from the sample convert  on average to one CCD count. To perform this calibration, the laser is tuned to the stop band of the sample around $770-780$ nm, i.e. in the range where the polariton emission occurs. The laser is reflected from a mirror placed at the position of the sample to simulate the sample emission. The power losses are recorded after each optical element in the detection path with a power meter and then correlated with a number of counts obtained on the CCD camera attached to the spectrometer. The calibration is performed including different linear polarisations of the laser light (changed with a $\lambda/2$ plate) to determine the mean collection efficiency of the spectrometer, because the diffraction gratings are polarization sensitive and the light emitted by low density polaritons is expected to be depolarized. Additionally, to rule out any background counts contribution to the calculated values, the efficiency is extracted from a slope of power-dependent series, by changing the duty cycle of the AOM (acousto-optical modulator). Using this procedure, we obtain a mean collection efficiency of the experimental setup.

The data analysis and extraction of the polariton density is performed based on the $k$-space (far-field) integration technique, which is a common approach to calculate the occupations of a given $k$-vector state in polariton research \cite{Kasprzak2006,Deng2006,Bhattacharya2013,SunBEC2017,Kammann2012}. The method is based on the accurate conversion of the counts collected at a given $k$-vector to an occupancy of this state, taking into account the experimental efficiency and the geometry of the detection. 

As mentioned in the main text, we filter out the trapped polaritons from the signal originating from the high-energy polaritons in the barrier by introducing a spatial filter of $\sim 10$ $\mu$m diameter in the image plane. The detected portion of the whole trapped polariton gas is scaled as the ratio of the spatial area of the filter to the effective area of the trap $A_{filter}/A_{trap}$ (assuming a uniform distribution within the trap). The analysed polariton dispersion is collected along the axis set by the monochromator entrance slit, which cuts through the middle of the circular symmetric far-field emission pattern. The mean photon emission rate at a given vector $k_i$ along the slit is related to the CCD counts as follows:
\begin{equation}
	 \frac{dN_{ph}(k_i)}{dt} = \eta \cdot I_{CCD}(k_i) \cdot \frac{A_{trap}}{A_{filter}},	
\end{equation}
where $i$ is the pixel position, $\eta$ is the collection efficiency and $I_{CCD}$ is the number of counts of the $i$th pixel.

The next step is to convert the photon emission rate to a mean number of polaritons occupying the $k_i$ state. In a steady state the mean photon emission rate is proportional to the number of polaritons, $N_{pol}$, multiplied by the polariton lifetime $\tau_{LP}$, which is determined by Hopfield coefficients:
\begin{equation}\label{eq:number}
	N_{pol}(k_i) = \frac{dN_{ph}(k_i)}{dt}\tau_{LP}(k_i),
\end{equation}

\begin{equation}\label{eq:lifetime}
	\frac{1}{\tau_{LP}(k_i)} = \frac{(1-|X(k_i)|^2)}{\tau_C}+\frac{|X(k_i)|^2}{\tau_X},
\end{equation}
where $\tau_C = 135$~ps \cite{Steger2013} is the cavity photon lifetime and $\tau_X$ is the exciton lifetime, being at least one order of magnitude larger, i.e. $\sim 1$ ns, in high quality GaAs/AlAs quantum wells \cite{Szczytko2004,Gurioli1991}. A common approach is to neglect the exciton decay rate contribution, which is much smaller than the photon decay rate. This approach becomes less relevant for highly excitonic polaritons $|X|^2>0.8$, where the excitonic contribution extends the polariton lifetime (\ref{eq:lifetime}), and discarding it would lead to overestimation of the polariton number (\ref{eq:number}). Hence, one can assume that this approach gives an estimate from above for the measured polariton density.

To evaluate the occupation $\mathcal{N}$of the $k_i$ vector, one has to calculate the number of states $N$ at this vector $k_i$:
\begin{equation}
	\mathcal{N}(k_i) = \frac{N_{pol}(k_i)}{N(k_i)}.
\end{equation}

\begin{figure}[t]
	\includegraphics[width=0.45 \textwidth]{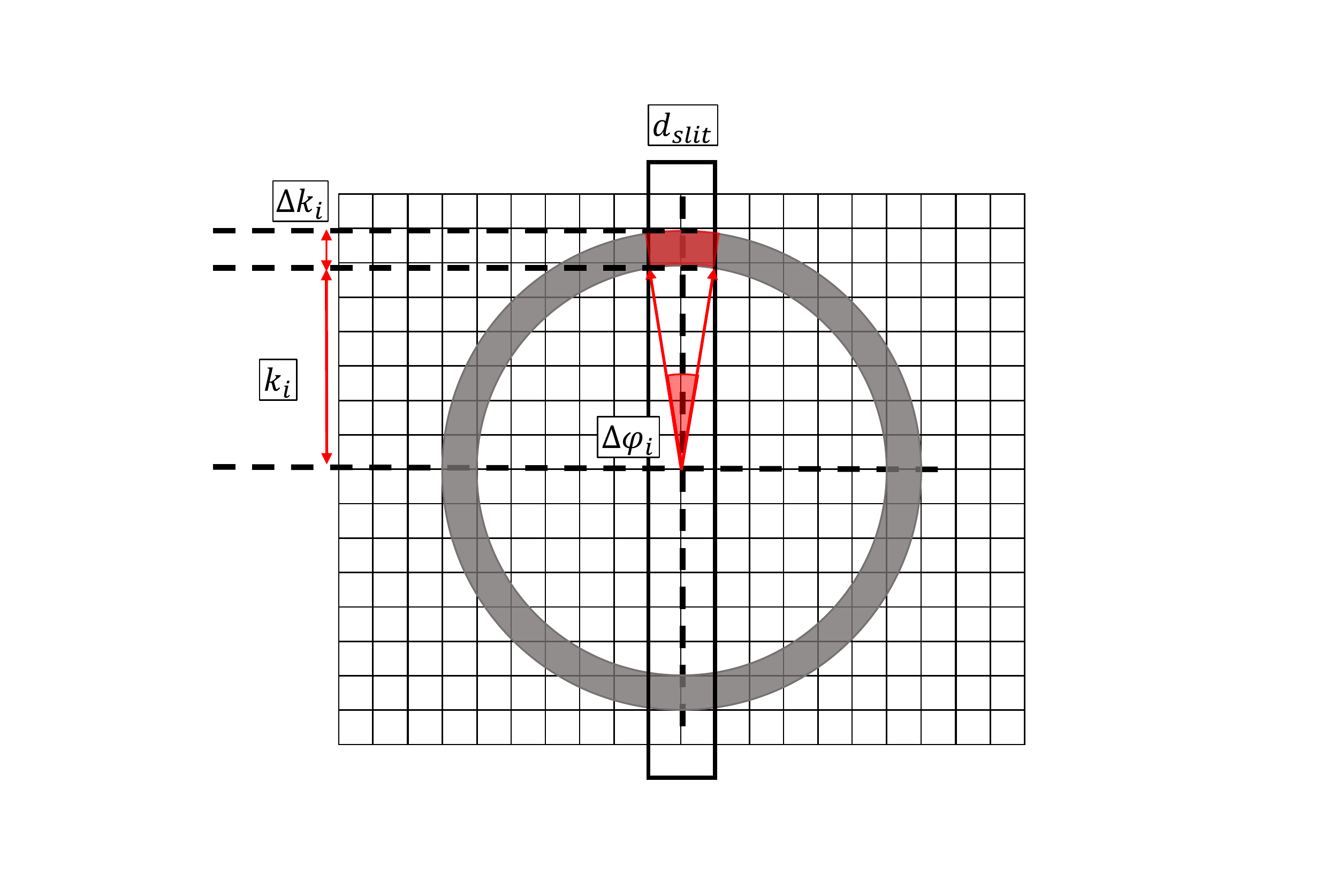}
\caption{\label{figA1} Schematics of a $k$-space integration geometry on top of the pixel array of the CCD camera showing the cut of an image with a spectrometer entrance slit. Symbols are explained in the text.}
\end{figure}

The number of states is determined by the geometry of our detection (see Fig. \ref{figA1}). The states at the $k_i$ position are selected by the width of one pixel $\Delta k_i$ in the $y$ direction and the width of the monochromator slit in the k-space plane. In cylindrical coordinates it yields the subtended k-state area $d\Omega(k_i) = k_i \Delta k_i \Delta \varphi_i$, where the angle arc measure is $\Delta \varphi_i = d_{slit}/R_i$ in the physical dimensions ($d_{slit}$ is the slit width and $R_i$ is the radius of the $k_i$ ring). Taking into account the volume of a single state, we obtain:
\begin{equation}
	N(k_i) = \Gamma\frac{k_i \frac{d_{slit}}{R_i}\Delta k_i}{\frac{(2\pi)^2}{A_{trap}}},
\end{equation}
where $\Gamma=2$ stands for the polariton state spin degeneracy. The excitation laser was chopped with an AOM with a duty cycle $d = 10 \%$, so the recorded number of polaritons has to be corrected by this factor. The final expression for the experimentally determined mean number of polaritons per state is as follows,
\begin{equation}
	\mathcal{N}(k_i) = \frac{4\pi^2}{k_i \frac{d_{slit}}{R_i}\Delta k_i \Gamma d} \frac{\eta}{A_{filter}} I_{CCD}(k_i)\tau(k_i).
\end{equation}

The final part of the analysis is to calculate the mean density of the polariton gas within the photo-generated trap. To obtain the total number of polaritons in the trap $N_{tot}$, one has to integrate the full experimental $k$-space. As we recorded only a central slice of the full far-field, it has to be integrated angularly under an assumption of the symmetry of the distribution in far-field. The final polariton density is expressed as:
\begin{eqnarray}
	n &=& \frac{N_{tot}}{A_{trap}} = \frac{1}{A_{trap}} \int_{-k_{max}}^{k_{max}} \pi k\left(\frac{4\pi^2}{A_{trap}}\right)^{-1} \mathcal{N}(k) dk  \\ \nonumber
	&=& \int_{-k_{max}}^{k_{max}} \frac{k}{4\pi} \mathcal{N}(k) dk.
\end{eqnarray}
Here we do not assume any distribution for polaritons, as the thermalised Bose-Einstein distribution is often not the case for polaritons.

The method used in our approach aims to minimize the number of assumptions, however several of them that we do make are not valid under all circumstances. As discussed in the main text, the linear slope of the effective potential due to the cavity wedge influences the shape of the ground state, so the assumption of uniform distribution of polaritons in the spatial signal collection area, as well as that of cylindrical symmetry of the far-field emission are poorly satisfied for large-area traps at very low pump powers. Finally, the limitation of the numerical aperture of the optical system prevents us from collecting all of the optically active polaritons in the field of view, which decreases the detected fraction of polaritons.

\begin{figure}
	\includegraphics[scale = 0.125]{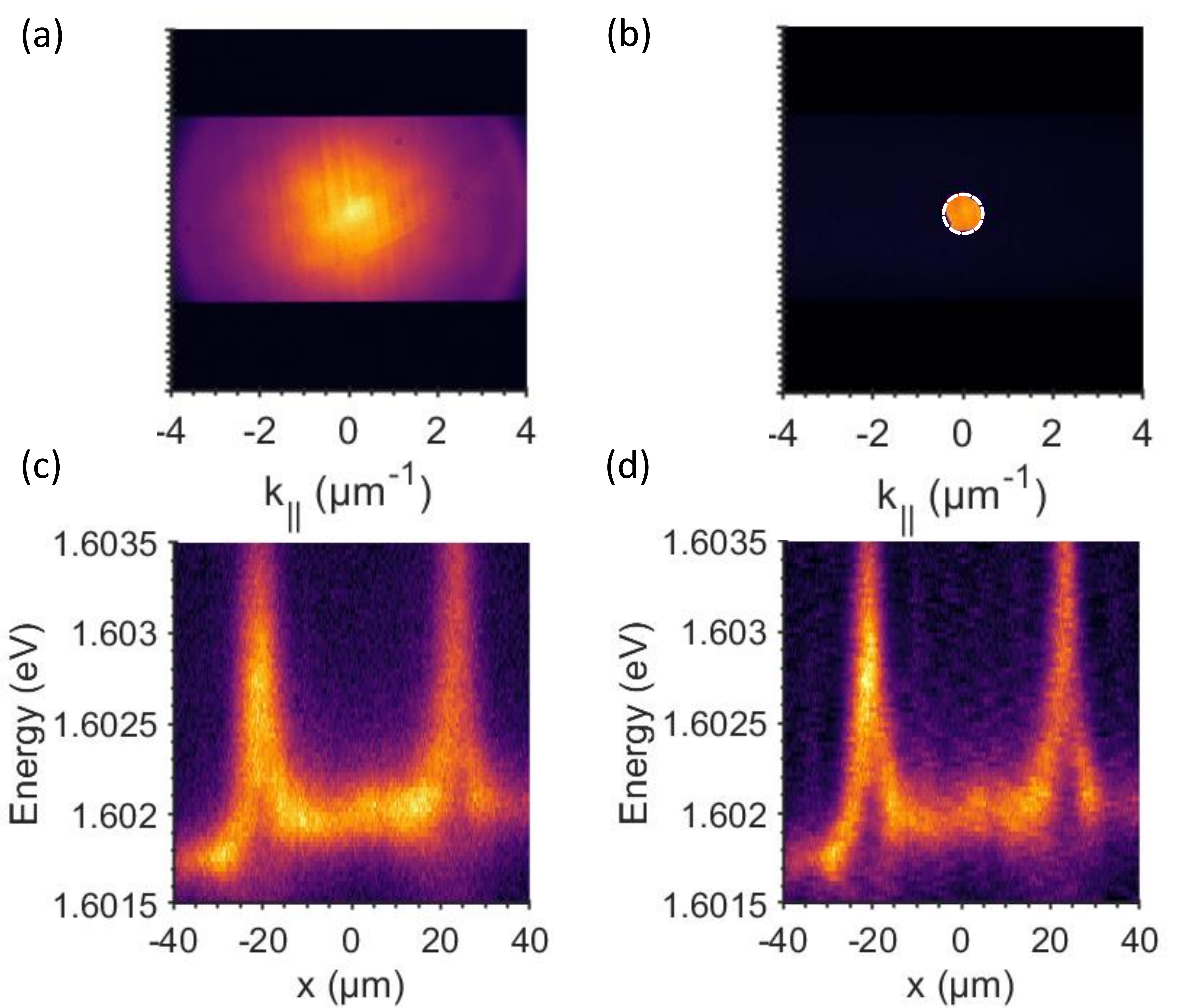}
\caption{\label{figA2} (a) Full far-field spectrum of low density polaritons at excitonic detuning. (b) Recorded position and size of the ${\bf k}_{\parallel} \approx 0$ filter. (c) Recorded spectrum with the use of momentum filtering in the case of a large excitonic trap. (d) The same as in (c) after using the Lucy-Richardson deconvolution. }
\end{figure}

\section*{Appendix B: Extraction of the local potential energy}

The extraction of the potential energy in the case of excitonic detuning was performed using the far-field filtering technique. We introduced an additional optical iris in the far-field (k-space) image plane, as depicted in Fig.~\ref{figA2}(a,b). According to the discussion in Sec.~\ref{sec:discussion}, it results in broadening of the image in the conjugate real-space plane due to diffraction on this filter. We recorded the point-spread function, which measures the broadening of the diffraction limited focused laser spot after imaging it with inclusion of a momentum filter, and obtained a value of about 6~$\mu$m. This broadening is visible in the real-space spectra of the optical potential, shown in Fig.~\ref{figA2}(c). To minimise the effect of the filter-induced broadening, we treated the data numerically by deconvolving the image in the position axis with a Lucy-Richardson algorithm, taking into account a Gaussian kernel of measured width. The effect of deconvolution is presented in Fig.~\ref{figA2}(d), where one can see sharpening of the blurry image of Fig.~\ref{figA2}(c) without additional distortion. Further, the extraction of the potential was performed by fitting the spectrum at each spatial position with a Lorentzian function to extract the $E_{LP}({\bf k_{\parallel}}=0,{\bf r})$.

\bibliographystyle{apsrev4-1}
\bibliography{References}

\end{document}